\documentclass[11pt,fleqn]{article}
\usepackage{amssymb}

\usepackage{amsmath}

\usepackage{amsmath,amssymb,amsthm}

\begin{document}
{\Large
\begin{center}
{\Large \textbf{General Results on Conditional Symmetry for the
Two-Dimensional Nonlinear Wave Equation}}

\vskip 20pt {\large \textbf{Irina YEHORCHENKO}}

\vskip 20pt {Institute of Mathematics of NAS Ukraine, 3
Tereshchenkivs'ka Str., 01601 Kyiv-4, Ukraine} E-mail:
iyegorch@imath.kiev.ua
\end{center}

\vskip 50pt
\begin{abstract}
We present full classification of $Q$-conditional symmetries for
the two-dimensional nonlinear wave equation
$$u_{tt} - u_{xx} = F(t,x,u)$$.
\end{abstract}

\newpage

\section{Introduction}
We discuss conditional symmetries of the two-dimensional nonlinear
wave equation
\begin{equation} \label{CS NLWE:Fwave}
u_{tt} - u_{xx} = F(t,x,u)
\end{equation}
for the real-valued function $u = u(t,x)$, $t$ is the time
variable, $x$ is the space variable. Further we will use the
following designations for the partial derivatives:
\[
u_t = - \frac{\partial u}{\partial t}; \\ u_x = - \frac{\partial
u}{\partial x}; \\ u_{tt}=\frac{\partial^2 u}{\partial t^2}; \\
u_{xt}=u_{tx}=\frac{\partial^2 u}{\partial t \partial x}; \\
u_{xx}=\frac{\partial^2 u}{\partial x^2}.
\]

The general equation in the class (\ref{CS NLWE:Fwave}) is not
invariant with respect to any operators, with only particular
cases having wide symmetry algebras, see e.g. \cite{CS
NLWE:FSerov-dA}.

The maximal invariance algebra of the equation (\ref{CS
NLWE:Fwave}) with $F=F(u)$ (not depending on $t,x$ and not equal
to zero) is the Poincar\'e algebra $AP(1,1)$ with the basis
operators
\begin{gather*}
 p_t  =  \frac {\partial}{\partial t}, \\ p_x  =  \frac {\partial}{\partial x},
 \qquad J = t p_x + x p_t.
\end{gather*}

The invariance algebras of the equation (\ref{CS NLWE:Fwave}) will
also include dilation operators for $F=\lambda u^k$ or $F=\lambda
\exp u$.

The symmetry of the linear equation (\ref{CS NLWE:Fwave}) with
$F=0$ and $F=\lambda \exp u$ is infinite-dimensional.

Similarity solutions for the equation (\ref{CS NLWE:Fwave}) can
be found by symmetry reduction with respect to non-equivalent
subalgebras of its invariance algebras
\cite{{CS NLWE:FSerov-dA},{CS NLWE:Tajiri84},{CS NLWE:FSS},
{CS NLWE:FbarAsolDA90}}.

Here we present the general result on the  $Q$-conditional
invariance of the equation (\ref{CS NLWE:Fwave}).

\newpage

\section{Conditional symmetry}

The concept of conditional symmetry in general (additional
invariance under arbitrary additional condition) and a narrower
concept of the  $Q$-conditional invariance (the additional
condition has the form $Qu=0$) was initiated and discussed in the
papers \cite{CS NLWE:OlverRosenau,CS NLWE:FTsyfra,CS
NLWE:FZhdanovCond, CS NLWE:ClarksonKruskal,CS NLWE:LeviWinternitz}
and later it was developed by numerous authors into the theory and
a number of algorithms for studying symmetry properties of
equations of mathematical physics. The importance of investigation
of the  $Q$-conditional symmetry was presented (see e.g.\cite{CS
NLWE:zhdanov&tsyfra&popovych99}) where equivalence of the
$Q$-conditional invariance and reducibility of the equations by
means of ansatzes determined by such operators $Q$ was proved.

Here we will use the following definition of the $Q$-conditional
symmetry:

\vskip 3pt

\noindent {\bf Definition 1.} {\it The equation
$\Phi(x,u,\underset{1}{u},\ldots , \underset{l}{u})=0$, where
$\underset{k}{u}$ is the set of all $k$th-order partial
derivatives of the function $u=(u^1,u^2,\ldots ,u^m)$, is called
$Q$-conditionally invariant \cite{CS NLWE:FSS} under the operator
\[
Q=\xi ^i(x,u)\partial_{x_i}+\eta ^r(x,u)\partial_{u^r}\nonumber
\]
if there is an additional condition
\begin{equation}
Qu=0, \label{CS NLWE:G=0}
\end{equation}
such that the system of two equations $\Phi=0$, $Qu=O$ is
invariant under the operator $Q$}. All differential consequences
of the condition $Qu=0$ shall be taken into account up to the
order $l-1$.

This definition of the conditional invariance of some equation is
based on what is in reality a Lie symmetry (see e.g. the classical
texts \cite{{CS NLWE:Ovs-eng},{CS NLWE:Olver1},{CS
NLWE:BlumanKumeiBook}}) of the same equation with a certain
additional condition. Conditional symmetries of the
multidimensional nonlinear wave equations are specifically
discussed in \cite{{CS NLWE:FSerov88},{CS NLWE:BarMosk},{CS
NLWE:zhdanov-panchakBoxu}}.

\newpage

\section{Previous papers on the problem}

Solving of the particular problem we discuss here was started by
P.Clarkson and E. Mansfield in \cite{CS NLWE:Clarkson Mansfield CS
NLWE} (the case $f=f(u)$), where the relevant determining
equations were written out but not solved, and continued by M.
Euler and N. Euler in \cite{CS NLWE:Euler CS NLWE}. In the latter
paper the determining conditions for the $Q$-conditional
invariance were taken without consideration of the differential
consequences of the condition $Qu=0$, so the resulting operators
did not actually present the solution.

Following \cite{CS NLWE:Clarkson Mansfield CS NLWE}, we will
consider the equation equivalent to (\ref{CS NLWE:Fwave}) of the
form
\begin{equation} \label{CS NLWE:yz-wave}
u_{yz} = f(y,z,u).
\end{equation}

The search for the operators of $Q$-conditional invariance in the
form
\begin{equation} \label{CS NLWE:Q-operator}
Q=a(y,z,u)\partial_y+b(y,z,u)\partial_z+c(y,z,u)\partial_u.
\end{equation}

\vskip 20pt

\section{Equivalence Transformations}
We will classify the systems of the type (\ref{CS NLWE:yz-wave}),
(\ref{CS NLWE:Q-operator}) under its equivalence transformations,
that is write down a set of all such systems that cannot be
transformed into each other by means of equivalence
transformations. However, here we will not be looking for
description of such group, but first limit ourselves by
equivalence up to some obvious transformations.

Further it will be expedient to look at equivalence
transformations for the special cases as they be different from
those in the general class.

The concept of equivalence of $Q$-conditional symmetries was
introduced by R. Popovych in \cite{CS NLWE:Popovych2000}

\newpage

\section{Determining Equations}
It is obvious that we can consider three inequivalent cases when
$a=0$, $b \ne 0$, then we can take
\begin{equation} \label{CS NLWE:Q-operator a=0}
Q=\partial_z+L(y,z,u)\partial_u
\end{equation}

and $a \ne 0$, $b \ne 0$, then we can take
\begin{equation} \label{CS NLWE:Q-operator a ne 0}
Q=\partial_y + K(y,z,u)\partial_z+L(y,z,u)\partial_u.
\end{equation}
where $K(y,z,u)\ne 0$.

The case $a \ne 0$, $b=0$ is equivalent to $a=0$, $b \ne 0$.

The case $a=0$, $b=0$ is trivial.

The additional condition $Qu=0$ will be represented respectively
by the equations
\begin{equation} \label{CS NLWE:Qu a=0}
u_z=L(y,z,u),
\end{equation}
and
\begin{equation} \label{CS NLWE:Qu a ne 0}
u_y+K(y,z,u)u_z=L(y,z,u).
\end{equation}

The determining equations have the form
\begin{equation}
-K_u^2+K_{uu}K=0, \label{CS NLWE:det eq1}
\end{equation}

\begin{equation}
-K L_{uu}+ \frac{K_u K_y}{K}+\frac{K_u^2 L}{K} +K_u(L_u-K_z)-
K_{uy}-L K_{uu}+KK_{zu}=0, \label{CS NLWE:det eq2}
\end{equation}

\begin{gather}
L_{uy}-L_{uz}K+L_{uu}L - \frac{L_u K_y}{K}+\frac{K_y K_z}{K} -
K_{yz} \nonumber \\
-3K_uf-\frac{K_u L}{K}(L_u-K_z)+K_u L_z -K_{zu}L =0, \label{CS
NLWE:det eq3}
\end{gather}

\begin{equation}
-f_y-Kf_z-Lf_u+L_{yz}+L_{uz}L+L_u f -
\frac{K_y}{K}(L_z-f)-K_zf-\frac{K_u L}{K}(L_z-f)=0 \label{CS
NLWE:det eq4}
\end{equation}

\newpage

\section{Main Results}
{\bf Case 1}. $K_u=0$, $K \ne 0$. The determining equations have
the form
\begin{equation}
-K L_{uu}=0, \nonumber
\end{equation}

\begin{equation}
L_{uy}-L_{uz}K+L_{uu}L - \frac{L_u K_y}{K}+\frac{K_y K_z}{K} -
K_{yz}  =0, \nonumber
\end{equation}

\begin{equation}
-f_y-Kf_z-Lf_u+L_{yz}+L_{uz}L+L_u f - \frac{K_y}{K}(L_z-f)-K_zf=0
\nonumber
\end{equation}

We have $K=k(y,z)$, $L=s(y,z)u+d(y,z)$. Using equivalence
transformations we can put $d(y,z)=0$.

From the determining equations we get
\begin{gather}
k(y,z)=\frac{T_y}{T_z}, \nonumber \\
s(y,z)=\frac{T_{yz}}{T_z}, \nonumber
\end{gather}
where $T=T(y,z)$ is an arbitrary function.

In this case the ansatz reducing equation (\ref{CS NLWE:yz-wave})
will have the form
\begin{equation} \label{CS NLWE:ansatz}
u=\sigma(y,z)\phi(\omega),
\end{equation}

where $\omega=\omega(y,z)$ is a new variable,
\begin{gather}
T_y \omega_z+ T_z \omega_y=0, \nonumber \\
T_y \sigma_z+ T_z \sigma_y=\sigma T_{yz}. \nonumber
\end{gather}
A reduced equation will have the form:
\begin{equation} \label{CS NLWE:reduced}
\sigma_{yz}\phi + \phi'(\omega_y \sigma_z + \omega_z \sigma_y +
\sigma \omega_{yz}) +\phi''\sigma \omega_y \omega_z =f,
\end{equation}
where $f$ satisfies the relevant conditions (\ref{CS NLWE:det
eq4}).

\newpage

{\bf Case 2}. $K=0$ - the case is equivalent to $a=0$, with the
additional condition (\ref{CS NLWE:Qu a=0}). Here we have
equations
\begin{equation} \label{CS NLWE:q 2.1}
u_y=L, u_{yz}=f.
\end{equation}
 The determining equations have the form

\begin{equation}
L_{uy}+L_{uu}L =0, \nonumber
\end{equation}

\begin{equation}
-f_y-Lf_u+L_{yz}+L_{uz}L+L_u f =0 \nonumber
\end{equation}

This case is actually equivalent to a pair of first-order
equations
\begin{equation} \label{CS NLWE:f 2.1}
u_y=L, u_{z}=\frac{f-L_z}{L_u}.
\end{equation}

{\bf Case 3}. $K_u\ne 0$, then $K_{uu}K=K_u^2$,
$K=k(y,z)exp(l(y,z)u)$. We can put $k=1$ and prove from the
resulting determining conditions $l_y=l_z=0$, so we can put $l=1$.

Then we can found that $L=s(y,z)expu+d(y,z)$. It is possible to
reduce this case to $k=1$, and we get the following determining
equation for $f$ with arbitrary $s$ and $d$:

\begin{equation} \label{CS NLWE:f 3.1}
f=\frac{1}{3}(s_y+d_z),
\end{equation}
so $f$ in this case depends only on $y$ and $z$, and the equation
$u_{yz}=f(y,z)$ is equivalent to the equation $u_{yz}=0$.

The conditions for $s$ and $d$ have the form
\begin{equation}
2s_{yz}-sd_z+2s_ys-d_{zz}=0,\\ -s_{yy}+2d_{yz}+s_yd-2d_zd=0.
\end{equation}

\newpage

\section{Conclusions}
We have considered the equations
\begin{equation} \nonumber
u_{yz} = f(y,z,u)
\end{equation}

with $f$ depending on $y,z,u$.

For such class the only nontrivial case is Case 1, $K_u=0$, $K \ne
0$.

The case $f=f(u)$ required special consideration, and has more
inequivalent cases.

}
\end{document}